\documentclass[sigconf]{acmart}

\usepackage{amsmath}
\usepackage{graphicx}
\usepackage{here}
\PassOptionsToPackage{hyphens}{url}
\usepackage{hyperref} 
\usepackage{url} 
\usepackage{textcomp} 
\usepackage{multicol}
\usepackage{subcaption}
\usepackage{comment}
\usepackage{booktabs,dcolumn}
\usepackage{multirow}
\usepackage{xspace}

\usepackage{multirow}

\usepackage{microtype}

\AtBeginDocument{%
  }

\copyrightyear{2026}
\acmYear{2026}
\setcopyright{cc}
\setcctype{by}
\acmConference[DIS '26]{Designing Interactive Systems Conference}{June 13--17, 2026}{Singapore, Singapore}
\acmBooktitle{Designing Interactive Systems Conference (DIS '26), June 13--17, 2026, Singapore, Singapore}
\acmDOI{10.1145/3800645.3812997}
\acmISBN{979-8-4007-2563-0/2026/06}

\newcommand{\tool}{\texttt{Robo-Blocks}\xspace}
\newcommand{\systemname}{\texttt{Robo-Blocks}\xspace}

\begin{document}

\title{\tool: Generative Scaffolding in End-User Design and Programming of Social Robots}

\settopmatter{authorsperrow=4}

\author{Arissa J. Sato}
\orcid{0000-0002-1103-8050}
\authornotemark[1]
\affiliation{%
  \institution{Department of Computer Sciences\\University of Wisconsin--Madison}
  \country{} 
}
\email{asato@wisc.edu}

\author{Callie Y. Kim}
\orcid{0009-0001-4195-8317}
\authornote{Both authors contributed equally to this research.}
\affiliation{%
  \institution{Department of Computer Sciences\\University of Wisconsin--Madison}
  \country{} 
}
\email{cykim6@wisc.edu}

\author{Nathan Thomas White}
\orcid{0009-0000-9414-9647}
\affiliation{%
  \institution{Department of Computer Sciences\\University of Wisconsin--Madison}
  \country{} 
}
\email{ntwhite@wisc.edu}

\author{Abhinav Maneesh}
\orcid{0009-0009-3487-0892}
\affiliation{%
  \institution{Department of Computer Sciences\\University of Wisconsin--Madison}
  \country{} 
}
\email{maneesh@wisc.edu}

\author{Yuqing Wang}
\orcid{0009-0000-7040-384X}
\affiliation{%
  \institution{Department of Computer Sciences\\University of Wisconsin--Madison}
  \country{} 
}
\email{yuqing.wang312@gmail.com}

\author{Hui-Ru Ho}
\orcid{0009-0000-3701-2521}
\affiliation{%
  \institution{Department of Computer Sciences\\University of Wisconsin--Madison}
  \country{} 
}
\email{hho24@wisc.edu}

\author{Bilge Mutlu}
\orcid{0000-0002-9456-1495}
\affiliation{%
  \institution{Department of Computer Sciences\\University of Wisconsin--Madison}
  \country{} 
}
\email{bilge@cs.wisc.edu}

\renewcommand{\shortauthors}{Sato et al.}

\begin{abstract}

Programming social robots is challenging for novice robot programmers due to required expertise in planning, interaction design, and programming. While large language models (LLMs) hold significant promise through code generation from natural-language descriptions, they can obscure critical elements of programming and supplant designer intent, eventually resulting in over-reliance instead of developing programming skills. 
In this paper, we explore how LLM-based social-robot-programming tools can support novice robot programmers through a \textit{Research through Design (RtD)} process. We designed and prototyped \tool, a block-based programming environment that leverages LLMs to offer novice robot programmers \textit{generative scaffolding} through structured narratives that connect high-level ideas to executable robot behaviors. 
Through deployment with novices, we discovered emerging user personas and usage patterns for generative scaffolding and showed how this scaffolding shapes end-user design and programming strategies. We present design insights for the effective use of generative scaffolding and its integration into the practice of social-robot programming.

\end{abstract}

\ccsdesc[300]{Human-centered computing~Collaborative interaction}

\keywords{Robot, End-User Programming, End-User Design, Programming Interface, Large Language Models, Generative Scaffolding}
\begin{teaserfigure}
  \includegraphics[width=\textwidth]
  {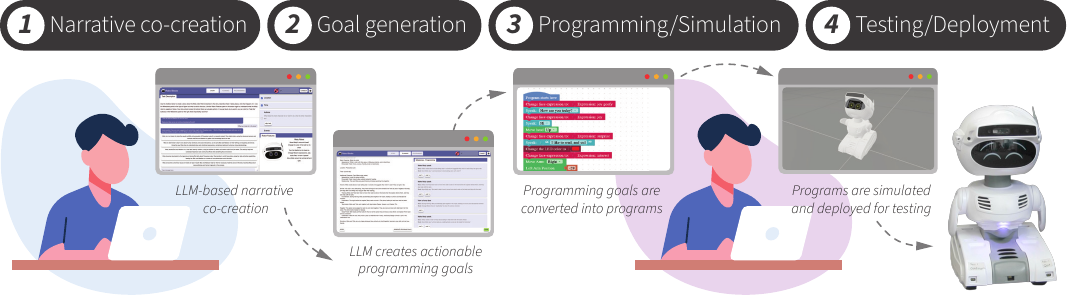}
  \caption{The generative scaffolding-based programming paradigm used in \tool{}. Users begin in (1) narrative creation, where they work with LLMs to create narratives featuring a robot. Then, users use LLMs in the (2) goal generation phase to create actionable programming goals from the narrative. These goals are then used in the (3) programming and simulation phase to build robot programs. Finally, the (4) testing and deployment phase has users run programs on a physical robot.}
  \Description{A horizontal four-phase flow diagram illustrating the generative scaffolding-based programming paradigm. Phase 1, labeled Narrative Co-creation, shows a user interacting with an LLM chat interface to create a robot narrative. An arrow leads to Phase 2, Goal Generation, where the LLM converts the narrative into a list of actionable programming goals. Another arrow leads to Phase 3, Programming and Simulation, where programming goals are converted into a block-based robot program shown on screen. A final arrow leads to Phase 4, Testing and Deployment, where the program is run on a physical Misty robot shown on the right. Each phase is connected by arrows indicating the sequential flow from narrative creation to physical robot deployment.}
  \label{fig:teaser}
\end{teaserfigure}


 \maketitle

\section{Introduction}

As robotic assistants become increasingly ubiquitous in daily environments, from service robots making deliveries at hotels, hospitals, and on the streets to domestic robots for cleaning, companionship, and children's education, there is a greater need to engage end users to design and program robot behaviors for these systems, tailoring how robots interact with people across diverse context through \textit{end-user design} and \textit{end-user programming}. 

However, designing and programming behaviors for robots to interact with people requires accounting for user intent during context-specific scenarios, anticipating how the robot should respond, and translating those responses into programmable tasks, requiring an integration of planning, interaction design, and programming, which are not skills that everyday users of social robots possess. We use \textit{end user} to refer to such individuals, specifically those without programming or robotics expertise who seek to author robot behaviors for social interaction scenarios~\cite{ajaykumar2021survey}. 

Recent research has proposed novel design methods to support end-user design, particularly conceptualizing and programming robot behaviors. One set of methods focuses on helping users model system behavior through embodied and contextual techniques, including bodystorming \cite{schleicher2010bodystorming}, which allows role-playing as robot agents \cite{porfirio2019bodystorming}, augmented reality to embed programming into real-world contexts \cite{kapinus2019spatially}, or tangible demonstrations of interaction between users and robots \cite{porfirio2021figaro}. These design methods aim to provide users with a relatable and understandable process for structuring and abstracting interaction to aid in interaction design and entirely automating programming. 

Another line of work simplifies the programming process through block-based languages, such as Blockly ~\cite{weintrop2017blockly},
Scratch~\cite{resnick2009scratch}, and OpenVP~\cite{schoen2024openvp}. 
Although these tools lack interaction design support, they allow users to specify interactive behaviors for the robot through exploration of drag-and-drop programming blocks, substantially lowering barriers to programming for novice users. 
The recent rise of large language models (LLMs) has motivated the use of natural language input to generate robot programs \cite{karli2024alchemist, mahadevan2024generative, mahadevan2025imageinthat, kim2024understanding, 10.1145/3654777.3676401, 10.1145/3757279.3785550}. 
However, while the use of LLMs further lower barriers to programming, the process of generating programs from natural language input lacks explicit consideration of interaction flow, user intent, and the role of context. 

Although these novel design methods and AI tools have helped to lower the barrier to entry, end-user design and programming remain challenging for novice robot programmers. They still need support in integrating planning, reasoning about interaction logic, and translating abstract ideas into executable behaviors. 
This difficulty highlights the need for methods and tools that effectively scaffold this process, particularly in supporting users as they conceptualize and plan human–robot interactions and translate them into programmable actions. Social robots are inherently character-driven and context-sensitive, designed to engage people through expressive behaviors, social cues, and adaptive interactions~\cite{BREAZEAL2003167}. Programming a social robot therefore requires reasoning about who the robot is, how it should express itself, and how it should respond to people across different situations, considerations that map naturally onto narrative elements such as character, emotion, action, and event. We propose that narrative-based generative scaffolding can bridge this gap, grounding the programming process in a medium that is intuitive for reasoning about social interaction.

In this paper, we engage in \textit{Research through Design (RtD)} to explore how natural-language-based approaches to capturing user intent can be integrated with block-based programming, which has been shown to substantially lower barriers in robot programming \cite{beschi2019capirci, huang2017code3, huang2016design, moros2019programming, ramouglu2017programming, weintrop2018evaluating}. Specifically, we build on the notion of \textit{generative scaffolding}---the use of generative AI tools as cognitive aids for users~\cite{Ma2025_ScaffoldingMetacognition, Liu2026-MetacognitiveScaffolding}---to provide end users with generated interaction narratives and programming goals as scaffolds for interaction design and programming tasks.

Our generative-scaffolding-based end-user design approach follows four phases in which users (1) first develop narratives that describe the robot's actions and behaviors; (2) translate these narratives into unit goals describing individual actions or changes in robot state; (3) build on these goals to explore, develop, and simulate a robot program; and (4) finally deploy and test their program on a physical robot (see Figure~\ref{fig:teaser}). 
We use this four-phase structure to make considerations of interaction flow, user adaptation, and the role of context explicit, while the LLM provides generative scaffolding by guiding users' transitions between phases.
This process contextualizes the conceptual stages of interaction design and programming within a narrative, while gradually introducing goal-oriented tasks for users to program. 
Within this paradigm, the LLM provides generative scaffolding by helping users create narratives, translating narratives into planning goals, mapping goals to robot actions, and enabling users to inspect, revise, and iterate on each intermediate representation.
Our work aims to answer the following research questions in the context of robot programming with novices: 
\begin{itemize}
    \item \textbf{RQ1.} \textit{How can generative scaffolding support end-user design and programming of social robot behaviors?}
    \item \textbf{RQ2.} \textit{What interactions, use patterns, and user perceptions emerge when using generative scaffolding?}
\end{itemize}

To address these questions, we designed and prototyped an end-user social robot design and programming environment, called \tool, that instantiated our four-phase process. We then conducted a user study that aimed to examine how generative scaffolding shapes narrative creation and programming outcomes, uncover the attitudes and use patterns that emerge in response to scaffolding, and identify design insights for bridging user intent and robot capabilities.
Our study provides empirical insight into how scaffolding shapes end-user design and programming, how novices interact with LLM suggestions, revealing distinct user personas and key usage patterns, how they perceived support through the interface, and how the juxtaposition of technologies---namely, \tool and the robot---shaped participants' expectations of both hardware and software capabilities.  

Our work makes the following contributions:
\begin{enumerate}
    \item A novel approach that supports end users in authoring social robot behaviors through narrative-based generative scaffolding;
    \item The design and development of an end-user design and programming environment, \tool, that instantiates our generative-scaffolding-based approach to end-user robot programming;
    \item An empirical understanding of how our approach facilitated end-user design and programming, including user personas and use patterns that reveal how novices engage with narratives and LLM suggestions during robot interaction design and programming.
\end{enumerate}



\section{Related Work}
Prior work has explored two complementary strategies for supporting novice robot programmers: narrative and scaffolding-based design methods that help users articulate and structure their intent, and technical tools that lower barriers to implementation. While each addresses part of the challenge, neither fully bridges the gap between early conceptual planning and executable robot behavior. We review each in turn below.

\subsection{Narrative and Scaffolding in End-User Design and Robot Programming}
End-user programming seeks to enable individuals to author robot behaviors without prior programming expertise~ \cite{ajaykumar2021survey}, and prior work has explored narratives and scaffolding strategies for bridging this gap, helping users articulate ideas and reason about interaction before engaging with programming tasks.

Narratives can help learners engage with the conceptual logic of a program before writing code \cite{parham2020does}. For example, \citet{yildiz2018digital} demonstrated how participants created storyboards and then translated them into Scratch programs. In this way, stories allow users to connect what they want to achieve conceptually with the corresponding programming tasks \cite{andersen2003teaching}. While narratives have been shown to enhance engagement and provide meaning, \citet{mcdermott2008more} emphasized the need for further investigation into the transitional processes by which novices develop expertise in algorithm construction.

Within robot programming, narrative-focused practices have been leveraged to support the design and development of robot behaviors. Bodystorming enables users to act out human-robot interactions through role-play, effectively capturing and translating human behaviors into robot programs \cite{porfirio2019bodystorming, abtahi2020presenting}. Similarly, other approaches capture user intent through narrative scenes, where figures, environments, and props provide immersive means for users to describe desired robot behaviors \cite{porfirio2021figaro, koike2024tangible}. 
Such practices encourage users to adopt the perspective of the robot and envision its actions through storytelling, thereby supporting the development of robot programs through narrative role-play. 
While these approaches emphasize enactment and demonstration, including interaction paradigms inspired by animal training~ \cite{10.1145/3610978.3640655}, they offer limited support for helping novices systematically translate narratives into programming artifacts. Our work addresses this gap by introducing narrative-based generative scaffolding that supports the systematic translation of user-created stories into programming goals and tasks.

\subsection{Technical Support for End-User Programming}
A variety of approaches have been explored to lower barriers in robot programming. These include augmented and virtual reality~\cite{yigitbas2021simplifying, ikeda2024programar, kapinus2019spatially}, visual programming environments ~\cite{schoen2020authr, pot2009choregraphe, huang2016design, buchina2016design, chowdhury2020user}, and block-based programming systems \cite{schoen2024openvp, fraser2015ten, resnick2009scratch}. Block-based programming has been shown to facilitate learning, particularly for novice programmers~\cite{mladenovic2018comparing}, and is often more accessible than text-based programming~\cite{weintrop2017comparing, weintrop2015block}. Leveraging these benefits, several tools have adapted block-based approaches specifically for robot programming~\cite{mayr2021considerations, weintrop2017blockly, bachiller2020learnblock}.
Another strategy for supporting end users involves embedding expert robotics knowledge directly into the system \cite{schoen2020authr, schoen2022coframe, sanders2018knowledge}.
By encapsulating this expertise, such systems can guide users in making program adjustments that account for the robot's capabilities or automatically correct errors. This form of support provides scaffolding to enhance end-user understanding \cite{janjanam2021design, buchanan1988fundamentals}, thereby lowering the barrier to entry by reducing the need for users to fully comprehend robotics before engaging with in programming tasks. 

More recently, LLMs have grown in popularity due to their capability to address a wide range of tasks. One application is in generating code and assisting the coding process \cite{kazemitabaar2023novices, fakhoury2024llm, liang2023understanding, 10.1145/3613904.3642773, 10.1145/3706598.3713748, 10.1145/3706598.3714154}. One commercial tool has already been released, GitHub Copilot, which is a tool capable of making suggestions for programmers to act on \cite{wermelinger2023using}. While such tools are effective for simpler tasks, their performance on more complex problems often depends heavily on the quality of the prompt and the user's expertise \cite{dakhel2023github}. Not only code generation, but LLMs have also been used to assist programmers by explaining code \cite{nam2024using}. LLMs have also been applied to robot programming, assisting users in generating complete programs and motion plans \cite{vemprala2024chatgpt, karli2024alchemist}, as well as synthesizing robot behaviors \cite{mahadevan2024generative, kim2024understanding}. By enabling users to define robot behaviors through natural language, LLMs have the potential to lower the barrier to entry in robot programming \cite{hu2024deploying}. 

Despite their capabilities, LLMs are not without limitations. Recent work has shown that while LLMs perform reasonably well in text-based robot programming, they struggle to build complete solutions in block-based environments \cite{shu2024llms}. Many implementations suffer from hallucinations \cite{zhao2023survey} and can forget instructions due to capacity \cite{zhao2023survey, jin2024llm}. Due to this forgetfulness, it is important to maintain users within the programming loop to monitor and guide LLMs \cite{chen2023forgetful}. To address these challenges, recent work has explored using linear temporal logic \cite{rozier2011linear} as guardrails to constrain LLM outputs and ensure they remain aligned with user intent \cite{yang2024plug}.

Taken together, prior approaches highlight important advances in lowering barriers to robot programming, but they primarily emphasize implementation and automation rather than scaffolding the early conceptual and planning processes that novices often struggle with. Our work complements these efforts by extending support to these upstream stages, integrating LLMs with block-based programming to help users move from initial ideas toward executable robot behaviors.

\section{\systemname: From Narrative to Robot Program}
We adopted a \textit{Research through Design (RtD)} approach~\cite{10.1145/1240624.1240704, 10.1145/3434074.3444868} to explore how LLMs might be integrated into social-robot programming-tools in ways that support novice users without obscuring designer intent or supplanting learning. Rather than treating LLMs primarily as mechanisms for code generation, we approached them as materials, probing how generative AI could function as \textit{generative scaffolding} to help users articulate, structure, and translate interaction ideas into executable robot behaviors. 

Through prototyping, deployment, and reflection, we developed \systemname, a block-based end-user design and programming environment for social robots. \systemname served as a research artifact that allowed us to explore and reflect on how generative scaffolding shapes novice users' design strategies, expectations of robot capabilities, and patterns of reliance on AI assistance. Our design process was guided by the following research question:
\begin{quote}
\textbf{RQ1.} \textit{How can generative scaffolding support end-user design and programming of social robot behaviors?}
\end{quote}

$\systemname$ is designed for novice robot programmers seeking to author social robot behaviors for companion and social interaction scenarios, such as programming a robot to greet and encourage a user, tell a story with expressive gestures, or act as a study companion. To illustrate the intended process, consider a user who wants to program Misty, a robot, to act as a study buddy. They begin by co-creating a narrative with the LLM describing the robot greeting the user, asking how their study session is going, and responding with encouragement. The system generates programming goals such as ``Have Misty greet the user'' and ``Have Misty ask about the study session,'' each accompanied by hints suggesting which blocks to use, such as speech blocks, face expression blocks, arm gesture blocks, and how to parameterize them. The user then assembles these behaviors using programming blocks and deploys the resulting program to Misty.

Design decisions were grounded in prior work in interaction design, end-user programming, and human-robot interaction, and the resulting artifact was deployed to surface empirical insights about how generative scaffolding is taken up in practice. Reflection on these insights informed the framing of our empirical study and the design principles derived from our findings.

\subsection{Design Considerations}
Our design was guided by four interrelated considerations, informed by prior research in end-user design and end-user programming. Guided by these considerations, we structured \systemname around a four-phase flow that incrementally transitions users from expressive intent to executable robot behavior: \textit{narrative creation}, \textit{goal generation}, \textit{programming and simulation}, and \textit{testing and deployment}. These phases were designed to support revision and backtracking during use, enabling users to revisit and refine narratives, goals, and programs as their understanding of the task and robot capabilities evolved.

\paragraph{Narrative as an Entry Point for Design.}
We positioned narrative creation as the starting point for robot programming based on prior work demonstrating that narrative and scenario-based methods effectively support novices in reasoning about and specifying desired social robot behaviors~\cite{koike2024tangible, porfirio2019bodystorming, KoaySyrdalDautenhahnWalters+2020+66+85}. 
This is particularly fitting for social robots, which are designed to engage people through expressive behaviors and adaptive interactions, as reasoning about what a social robot should do is fundamentally a question of social context, character, and intent, precisely the elements that narrative captures~\cite{BREAZEAL2003167, FONG2003143}.
For novice users, narratives therefore provide a familiar medium for articulating intent, allowing them to focus on \textit{what} the robot should do and \textit{why}, before addressing \textit{how} those behaviors are operationalized. By grounding programming in narrative descriptions of robot actions and social interactions, we aimed to scaffold users' transition from high-level ideas to structured, goal-oriented robot programming.

\paragraph{Generative AI as Scaffolds for User Agency}
A key design goal as to ensure that LLM assistance augmented, rather than replaced, user decision making. Instead of autonomously generating complete programs, the LLM was designed to offer suggestions, prompts and translations that users could inspect, modify or reject. This reflects a view of LLMs as cognitive scaffolds that support reflection and sensemaking while preserving user agency and authorship \cite{huang2026narrativescaffoldingnarrativefirstframework}. Intermediate representations such as generative narratives and goals were intentionally exposed to make reasoning visible and to mitigate over-reliance on opaque automation \cite{SCAIFE1996185, 5613437}.

\paragraph{Supporting Iteration in Use}
We intentionally designed \systemname to support iterative exploration \textit{during use}. End-user design and programming are exploratory activities, particularly for novices whose understanding of robot capabilities evolves over time \cite{ahmed2003understanding, BALL01111994}. The system therefore allows users to move between phases, revises narratives or goals, and refine programs in response to simulation or deployment outcomes. This supports learning through experimentation and reflection.

\paragraph{Balancing Abstraction and Transparency}
Finally, we sought to balance abstraction with transparency by introducing intermediate representations that bridge expressive intent and executable code. Goals and programming blocks served as concrete reference points connecting narrative elements to robot actions, allowing users to trace how high-level ideas are operationalized. By making these mappings visible, the system supports users in developing a mental model of how interaction concepts translate into robot behavior, rather than concealing this process with automated code generation \cite{andrews2023role, 5613437, CARROLL198845}.

\subsection{Narrative Creation} \label{sec:narrative-creation}
\begin{figure*}
    \centering
    \includegraphics[width=\linewidth]{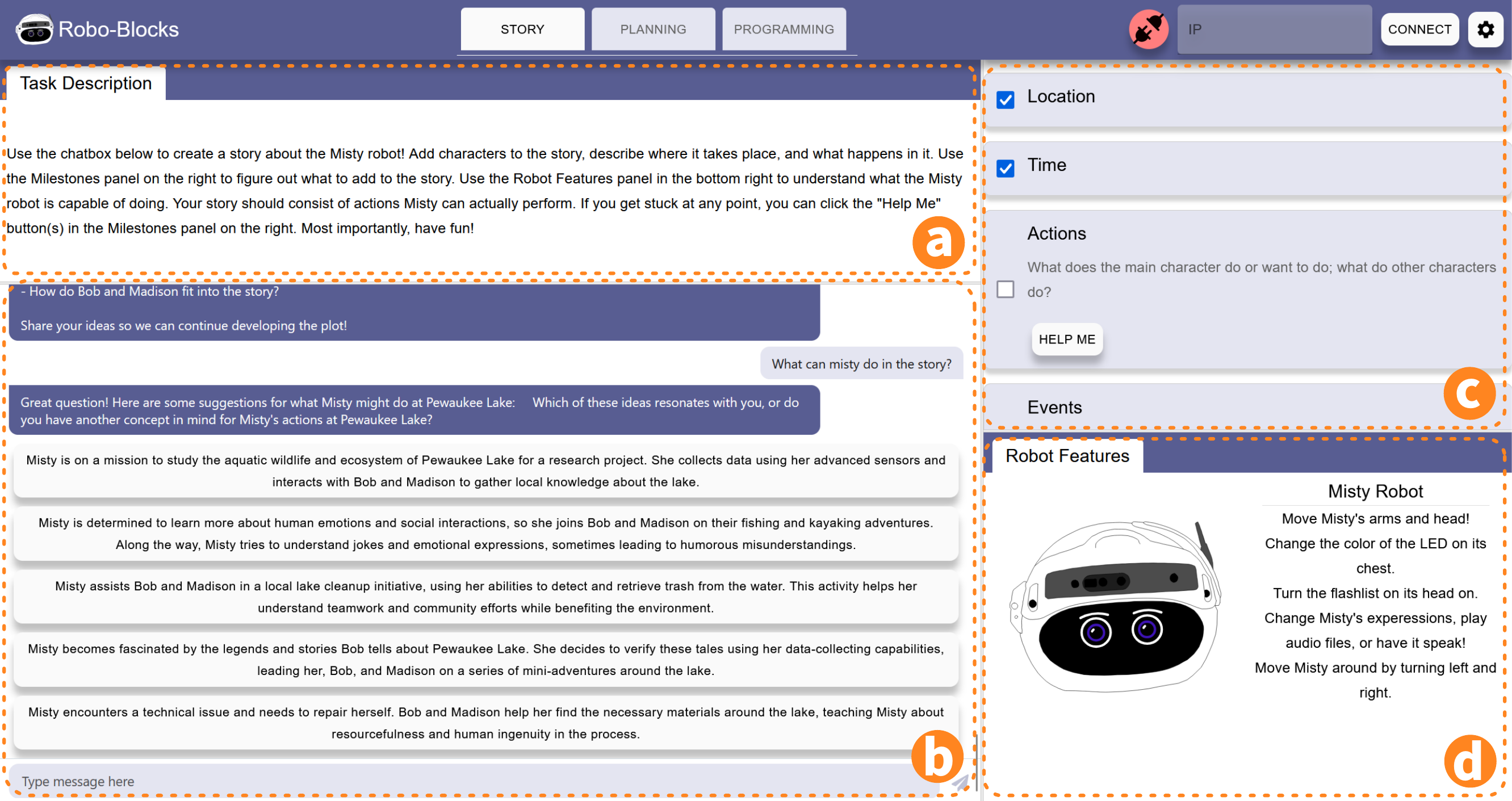}
    \caption{\systemname' interface for the narrative creation phase. Section (a) provides the user with a description of the task. Users can interact using the chat box in section (b) to interact with the LLM agent to create a narrative, using suggestions the agent provides. The user can then mark milestones complete in section (c) or request help from the LLM for a particular milestone. Section (d) provides a description of the robot's capability.}
    \Description{A screenshot of the system interface during the narrative creation phase, divided into four labeled sections. Section (a) on the upper left shows a task description panel with written instructions guiding the user on how to create a narrative. Section (b) in the center shows a chat interface where the user can type messages to interact with the LLM agent, with example suggestions displayed as clickable buttons above the text input. Section (c) on the right shows a milestone checklist with items such as Location, Time, Actions, and Events, each with a checkbox and a Help Me button that users can click to request LLM assistance for that milestone. Section (d) in the lower right shows a Robot Features panel displaying a photo of the Misty robot alongside a list of its capabilities, such as moving its arms and head, changing LED colors, and playing audio files.}
    \label{fig:narrative-creation}
\end{figure*}

The narrative creation phase is the first phase of the process in \systemname and reflects our design inquiry into how generative AI can scaffold early-stage planning without supplanting user intent. Rather than asking users to begin with commands or code, we invite them to articulate a narrative that describes the robot around the robot and its role within a situation, including its behaviors, actions, and emotions.

To support this process, we introduced a set of narrative milestones: \textit{characters}, \textit{locations}, \textit{time}, \textit{actions}, \textit{events}, \textit{ending}, and \textit{emotions}. These milestones provide structure, guiding users in what to consider during early-stage of planning (see Figure~\ref{fig:narrative-creation} (c)). 
The milestone structure was informed by Self-Regulated Strategy Development (SRSD) ~\cite{Mason2002_StoryMilestones}, an instructional framework that emphasizes breaking down open-ended tasks into smaller components while supporting novice learners in planning and self-monitoring. SRSD provided a structured categorization of storytelling elements that maps onto the components needed to describe robot behavior, making it crucial for scaffolding narrative creation in our system. This aligns with our goal of helping novice robot programmers structure their ideas during early-stage narrative construction, where novices may otherwise struggle with where to begin or how to organize their narratives.
Each milestone targets different aspects of a narrative and was chosen due to either being commonly found in storytelling (\textit{characters, locations, time, events, ending}) or playing a key role in defining robot behavior (\textit{actions, emotions}). 
Together, they prompt users to consider not only what happens in the story, but how the robot behaves and expresses itself across situations. By externalizing these considerations, the milestones make aspects of planning that are often implicit more explicit and available for reflection.

LLM support in this phase is deliberately designed as collaborative rather than fully automated by default (see our supplementary materials for the full prompt).\footnote{Supplementary materials can be found at \url{https://osf.io/eumc3/overview?view_only=e1c4d4b329e34330a14a1426bd95fef8}.} The system makes the user's process visible to the LLM (\textit{e.g.,} through milestone completion), enabling the model to provide context-specific suggestions rather than generic prompts. Users can request help at any milestone, receive multiple candidate suggestions, and selectively incorporate or ignore them (see Figure~\ref{fig:narrative-creation} (b)).
These methods of user interaction with the LLM allow for direct assistance with each of the milestones while leveraging the strength of LLMs for ideation. 

This phase functioned as a design probe to explore how structured interaction can shape the role of generative AI. By foregrounding user authorship and exposing AI suggestions selectively, the design allows us to examine how explicit generative scaffolding influences users' sense of control and engagement during end-user design for robot behavior.

\subsection{Translating Narrative into Programmable Goals}
\label{sec:goal-generation}
\begin{figure*}
    \centering
    \includegraphics[width=\linewidth]{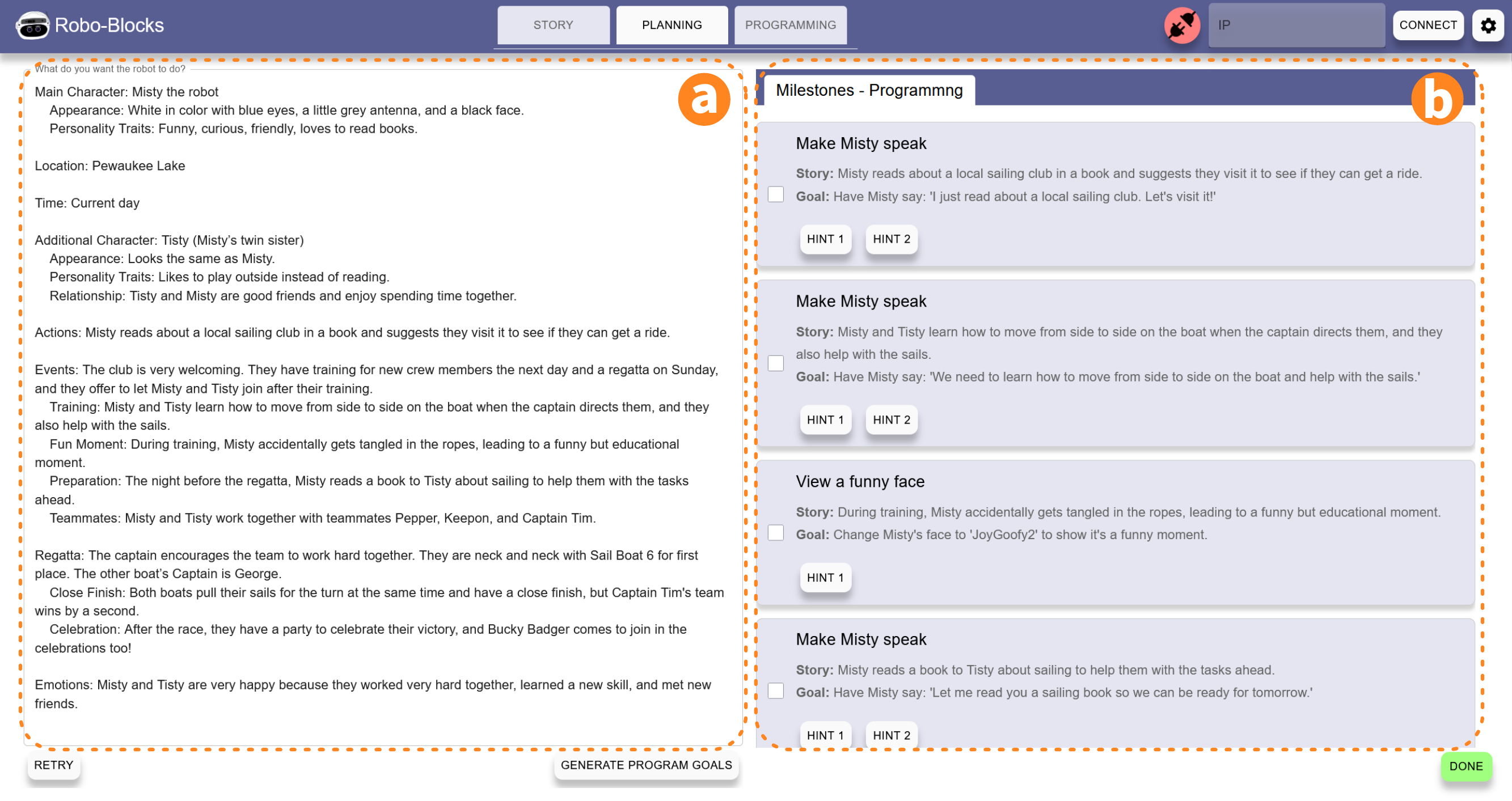}
    \caption{\systemname' interface for the goal generation phase. Users can use section (a) to provide their narrative to the LLM agent. The agent will then convert the story into programmable goals and display them in (b).}
    \Description{
    A screenshot of the system interface during the goal generation phase, divided into two labeled sections. Section (a) on the left shows a narrative input panel displaying the user's full narrative text, which describes the robot's story including characters, locations, actions, and events. Below the narrative text are three buttons: Retry, Generate Program Goals, and Done, allowing users to regenerate goals or proceed to the next phase. Section (b) on the right shows the generated programming goals panel, displaying a list of individual goals derived from the narrative. Each goal card shows the relevant story snippet it was derived from, the corresponding programming goal description such as ``Have Misty speak'' or ``View a funny face,'' and expandable hint buttons labeled Hint 1 and Hint 2 that users can click to reveal implementation suggestions. The two sections are displayed side by side, reinforcing the visual connection between the narrative content on the left and the derived programming goals on the right.
    }
    \label{fig:goal-gen}
\end{figure*}

The second phase explores how high-level narrative intent can be translated into programmable goals. In this phase, users collaborate with an LLM agent to decompose their narrative into a set of goals. Each goal is accompanied by optional hints that suggest possible implementation strategies and explicitly reference elements of the narrative (see Figure~\ref{fig:goal-gen} (b)).
By exposing this translation step (see Figure~\ref{fig:goal-gen}), \systemname makes the LLM's interpretation visible and understandable rather than implicit. This visibility allows users to reflect on and revise the mapping between intent and robot behavior. This is done either by revising the narrative, regenerating goals, or selectively adopting suggestions rather than passively accepting AI-generated outputs.

Hints serve as a secondary layer of scaffolding, further breaking down each programming goal into concrete steps for the user. Depending on the goal, hints may suggest which blocks to use, where to place them in the program, or how to parameterize them. Importantly, hints remain optional, allowing users to engage with the level of guidance that best supports their current understanding.

For this phase, we adopted a one-shot approach to the LLM agent, providing a single example of how a narrative can be translated into goals. 
The LLM prompt for this phase includes a description that its purpose is to transform user descriptions of robot behavior into concrete, actionable goals for programming. The prompt includes textual descriptions regarding the capabilities of the Misty robot, the locations of all programming blocks in the interface, and the parameters for each of the blocks and what they do. Within the prompt, a JSON schema is specified that the agent should use to return any output. Lastly, the agent is instructed that for each goal it creates, at least one hint should accompany it (see supplementary materials for the full prompt).

Through this phase, we examined how LLMs can support end-user design and reasoning rather than automate programming. By grounding goal generation in the robot's capabilities while keeping the narrative central, this phase supports a gradual transition from expressive intent toward implementation.

\subsection{Programming and Simulation}
\label{sec:programming-simulation}
\begin{figure*}
    \centering
    \includegraphics[width=\linewidth]{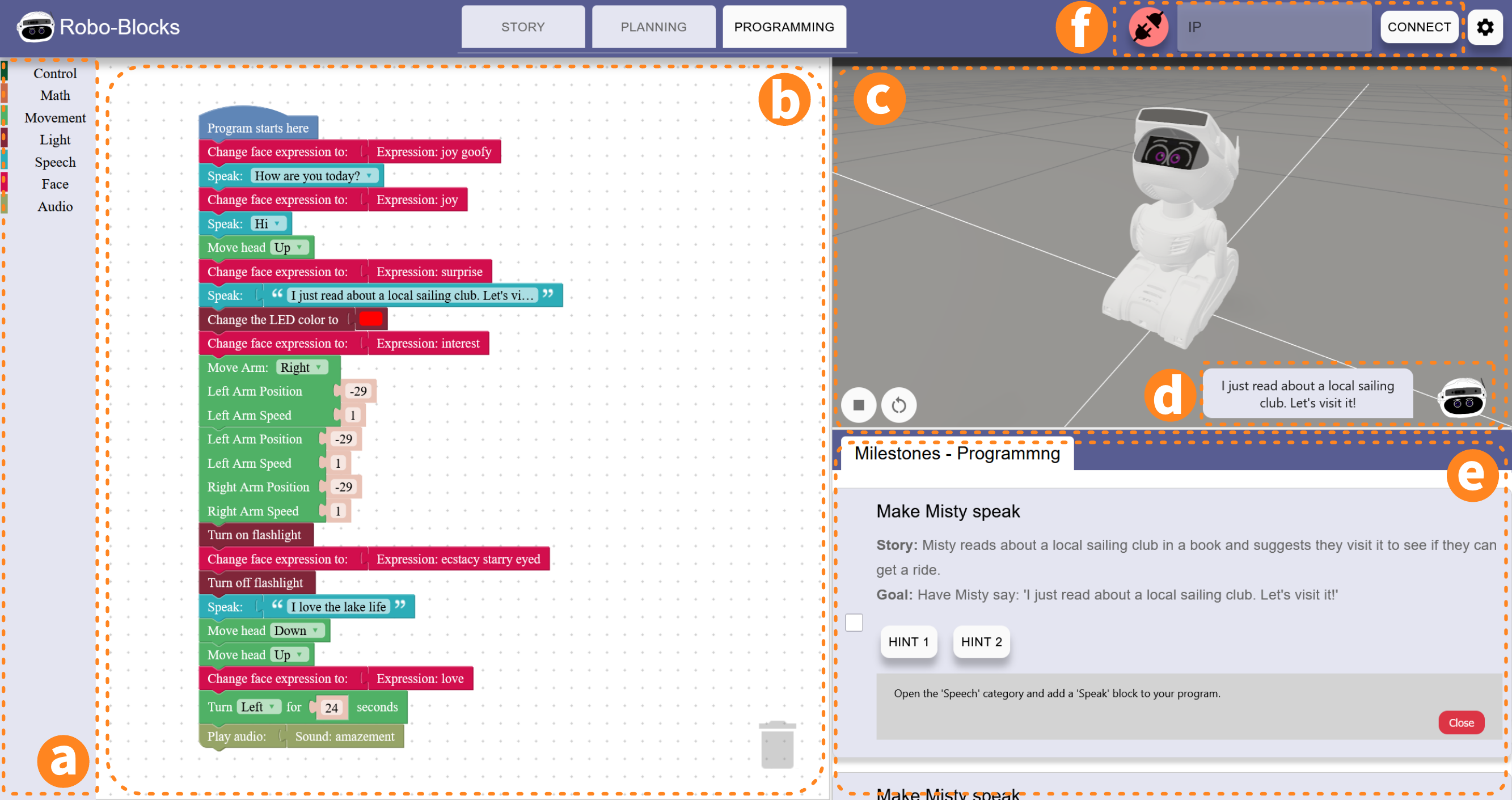}
    \caption{\systemname' interface for the programming and simulation and testing and deployment phase. Users can build their program in the Blockly canvas (b) using the blocks from the program drawer (a). Users can refer to their goals from the goal generation phase in (e), using the hints as needed to assist in building the program. At any time, users can run their program within the simulator (c) and the robot's programmed speech (d). When the user is ready to deploy the program on a physical robot, they can connect using (f) and deploy the program.}
    \Description{A screenshot of the system interface during the programming, simulation, and deployment phases, divided into six labeled sections. Section (a) on the far left shows a program drawer with categorized block groups including Control, Math, Movement, Light, Speech, Face, and Audio, which users can expand to access programming blocks. Section (b) in the center shows the Blockly programming canvas where users drag and drop blocks to build their robot program, with an example program visible showing blocks for changing face expressions, speaking, moving the head, and controlling lights. Section (c) on the upper right shows a built-in 3D simulator displaying a rendered model of the Misty robot, allowing users to visualize how their program will execute before deploying to the physical robot. Section (d) shows a speech display panel overlaid on the simulator that shows the text the robot will speak during program execution. Section (e) on the lower right shows the programming goals panel carried over from the goal generation phase, displaying individual goals with their associated story context, programming goal description, and expandable hint buttons labeled Hint 1 and Hint 2. Section (f) in the upper right corner shows a connection panel with an IP address input field and a Connect button that users can use to connect to and deploy their program to a physical Misty robot.
    }
    \label{fig:programming-simulation}
\end{figure*}
This phase examines how generative scaffolding can support programming through visibility, optional guidance, and reflection. Users construct programs using blocks that correspond to robot actions and expressions, guided by previously generated goals and hints.

Programming in \systemname takes place in a block-based environment built on Google's Blockly \cite{fraser2015ten}. 
Blockly enforces proper syntax relations by only allowing specific blocks to connect. 
This design simplifies the programming process by ensuring that users focus on the task without needing to learn syntax, worry about programming details, or deal with low-level implementation. 
Crucially, goals generated in the previous phase remain visible throughout programming (see Figure~\ref{fig:programming-simulation} (e)). This design supports reflection by allowing users to track progress and reason about how abstract intentions map into concrete implementations. Hints associated with each goal are optional and vary in specificity, supporting different levels of reliance and engagement for the user. 
To support reflection, \systemname includes a built-in simulator that visualizes robot motion, expressions, and speech. By allowing users to compare intended behavior (from goals) with observed behavior (from execution), the simulator functions as reflective feedback.

This phase was designed to examine how visibility and abstraction shape novice programming practices. By keeping goals, hints, and execution feedback simultaneously accessible, the system supports learning through construction and reflection while avoiding over-reliance often associated with AI code generation.

\subsection{Testing and Deployment}
\label{sec:testing-deployment}
While the built-in simulator provides users with a preliminary understanding of how programmed narrative might be executed by the robot, deployment on the physical robot remains a necessary step for understanding how planned interactions translate into real-world context.
The final phase of the programming process allows users to evaluate real-world program performance on the robot, ensuring that the robot's capabilities align with both the intended narrative and program, and allows users to make any final adjustments as needed.
This also allows users to verify if the output aligns with the simulation under real-world conditions. 

Rather than treating deployment as an endpoint, we designed it as part of an iterative process, encouraging users to revisit earlier phases when misalignments arise.
\systemname  supports this phase by allowing users to connect directly to a physical robot through the interface wit an entry field where users can input the IP address of the Misty robot and connect to it (see Figure~\ref{fig:programming-simulation} (f)). 
When connected, the connection icon will turn from red to green, and the unconnected cord will become connected, informing the user of a successful connection to the robot. 
After connecting to the robot, an additional toggle is placed on the simulator that allows the user to specify whether running the program occurs on both the simulator and the robot, or just on the simulator (see Figure~\ref{fig:programming-simulation} (c)). By requiring the simulator to be enabled when the program runs, the design allows users to verify whether the simulation and the real-world robot behavior align. This reinforces deployment as an opportunity for building expectations about robot behavior.

\subsection{Implementation}
We designed \systemname
around the Misty II robot\footnote{\url{https://www.mistyrobotics.com/misty-ii}} due to its low cost and easy integration with web applications. 
These factors make Misty a good candidate for rapid deployments and supported our goal of examining novice end-user design practices.
\systemname was built using React, the Misty II REST API, and gpt-4o \footnote{\url{https://openai.com/index/hello-gpt-4o/}} from OpenAI through Microsoft Azure\footnote{\url{https://azure.microsoft.com/en-us/products/ai-services/openai-service}}. 

\section{Exploring Use of Generative Scaffolding}

To explore the use of generative scaffolding in end-user design and end-user robot programming, we deployed \systemname with individual novice robot programmers.
This study enabled us to closely examine how generative scaffolding affects planning and programming, revealing distinct use patterns and interactions.
We also collected participants' reflections to better understand their perceptions of generative scaffolding.


\subsection{{Task and Procedure}}
Each user study session took approximately 1.5 -- 2.5 hours. Each session was facilitated by one to two experimenters who provided assistance (\textit{e.g.,} explained how to use the interface, clarified the exercise) to the individual as necessary, but did not guide participants' design decisions. Each participant was provided a computer station that consisted of a desktop computer, one keyboard, and one mouse/trackpad. 
Participants had their computer screens recorded to be used for later verification on how they used the system.
The following user study procedure was approved by the authors' Institutional Review Board (IRB). Materials used in this session can be found in the supplementary materials.

After confirming consent from participants, the experimenter introduced the purpose of the user study and activities.
The experimenter played a pre-recorded video that provided an introductory lecture on social robotics and a brief description of how to create a storyboard for an interaction, lasting for approximately 10 minutes. 
Following the lecture, the experimenter introduced the Misty robot and demonstrated some of its capabilities, through Misty's native user interface, for the participants' reference. This introduction phase lasted for approximately 15 minutes in total. 
Participants were then given 10 minutes to create a storyboard of a social interaction with the Misty robot based on the introductory lecture and demonstration without any use of online resources. 

\begin{figure}[t]
    \centering
    \includegraphics[width=\linewidth]{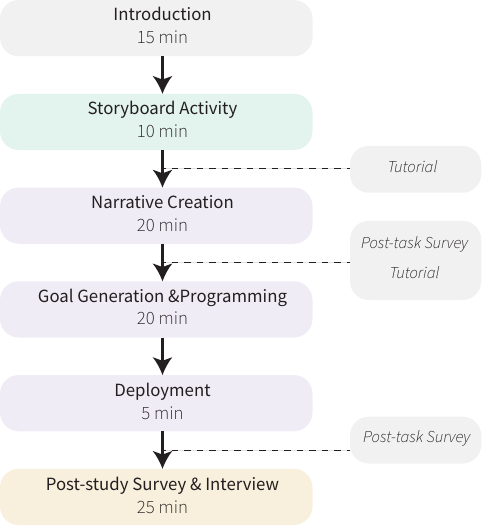}
    \caption{Study procedure of \systemname. Participants completed a storyboard activity, then engaged with four phases of \systemname: narrative creation, goal generation, programming, and deployment, each preceded by a tutorial video. Post-task surveys were completed after narrative creation and after deployment, followed by a post-study survey and semi-structured interview.}
    \Description{
    A vertical flowchart illustrating the study procedure. The flowchart begins at the top with an Introduction box labeled 15 minutes. An arrow leads down to a Storyboard Activity box labeled 10 minutes, where participants created a paper storyboard of a social robot interaction. An arrow leads to a Narrative Creation box labeled 20 minutes, preceded by a Tutorial label on the right indicating participants watched a tutorial video before this phase. A Post-task Survey label on the right indicates participants completed a survey after this phase. An arrow leads to a Goal Generation and Programming box labeled 20 minutes, also preceded by a Tutorial label on the right. An arrow leads to a Deployment box labeled 5 minutes. A Post-task Survey label on the right indicates participants completed a second survey after deployment. A final arrow leads to a Post-study Survey and Interview box labeled 25 minutes at the bottom of the flowchart. Tutorial videos and post-task surveys are shown as annotations on the right side of the flowchart, connected to their respective phases by horizontal lines.
    }
    \label{fig:study-design}
\end{figure}

Participants next completed two exercises using \systemname. Before the first exercise, participants watched a tutorial video introducing the narrative creation phase and its functions. Participants then used the narrative creation phase to co-author a story describing their intended social robot interaction (20 minutes), followed by a post-task survey asking about their perception and experience working with an AI agent. Before the second exercise, participants watched a tutorial video introducing the goal generation and programming phase. Participants then used goal generation and programming to translate their narrative into an executable robot program, then deployed it and observed it execute on the Misty robot (20 minutes). Because \systemname was designed to support iteration, participants could return to adjust generated goals before finalizing their program. Thus a second post-task survey covering both steps was administered after deployment. The session concluded with a post-study survey and a semi-structured interview, approximately 25 minutes combined (See Figure ~\ref{fig:study-design}).

\subsection{Participants}
We recruited $14$ participants (6 women, 8 men; age $M= 30.79$, $SD = 16.81$), all of whom self-reported to be of at most beginner-level proficiency in programming to better understand the novice programmer perspective.
We stopped recruitment after $14$ participants as we saw saturation in survey and interview responses. 
All participants were provided information about the goal of the study and were given the option to volunteer to participate in the study. 
Each participant was compensated at a rate of \$15 USD per hour for their time.
Details about the participants can be found in the supplementary materials.

\subsection{Data Collection}

\paragraph{Storyboard} Participants were first asked to create a storyboard of a social interaction with Misty, illustrating their ideas on paper without the aid of generative AI tools.
These storyboards served three purposes: 1) for the participant to brainstorm a narrative, 2) for later use in the interviews to prompt participants' reflections on creating narratives with and without AI support, and  3) for our analysis of the narrative creation process. 

\paragraph{Surveys} Prior to joining the study, participants were provided a screening survey to confirm they were of at most beginner proficiency in programming to align with the goals of this work to understand novice programmers. 
The screening survey also collected participants' proficiency in robotics and use of generative AI tools.
During the study, the participants completed a post-task survey to share their experience working with an AI agent, hereinafter referred to as an LLM agent, after each phase and describe any instances of conflict they may have experienced with the LLM agent.
Finally, the participants were provided a post-study survey, consisting of the SUS~\cite{brooke1996sus} and USE survey \cite{lund2001measuring} to understand their perception of using \systemname.  
The provided materials can be found in the supplementary materials. 

\paragraph{Screen Recording}
During the activities, participants' screens were recorded using OBS Studio\footnote{https://obsproject.com/} or Zoom\footnote{https://www.zoom.com/} to capture their interactions with \systemname.

\paragraph{Activity Outcomes}
A history of the participants' chat communication with the LLM agent was logged during the story creation, planning, and programming tasks.
In addition, we recorded an activity log of participant interactions with \systemname for later analysis and verification.
Finally, we collected the final programs produced during the programming activity.

\paragraph{Interviews}
Four authors conducted semi-structured interviews with individual participants about their experience using \systemname, lasting between $5$ and $16$ minutes.
Please refer to the supplementary materials for more details.

\subsection{Analysis}

Our analysis drew on multiple sources of data: participant storyboards, chat logs, activity logs, final programs, SUS~\cite{brooke1996sus} and USE~\cite{lund2001measuring} questionnaire results,  survey responses, interview transcripts, and screen recordings of each individual's interaction with \systemname. 

To examine how generative scaffolding supported end-user design and programming (RQ1), we analyzed each storyboard and the resulting story summary, focusing on the \textit{context} of the story, the \textit{character(s)} introduced, and the \textit{actions} designated for the Misty robot. 
This allowed us to characterize the core components of each narrative as well as major changes in the storyline or details.
Two authors independently extracted these features and cross-checked each other's coding to ensure fidelity to the participants' original storyboards and story summaries.
We also reviewed screen recordings to capture participants' behaviors, noting moments when they accepted and rejected suggestions from \systemname.

To identify interactions, use patterns, and user perceptions (RQ2), we examined the activity logs and screen recordings to trace how participants engaged with the system. 
Specifically, we identified instances of hint usage and how hints were incorporated into the final program. 
By triangulating data from logs, programs, storyboards, and observations, we identified emerging use patterns. 
For the narrative creation phase, this included frequency of help requests, acceptance of suggestions from \systemname, analysis of storyboard and narrative creation outcomes, and observations from screen recordings. 
For the programming phase, we included frequency and usage of hints, analysis of the final program, and observations from screen recordings.

To understand how novices perceive and respond to generative scaffolding, we analyzed SUS and USE questionnaire results, survey responses, and interview transcripts. 
For SUS and USE questionnaires, we averaged responses across subscales and calculated means and standard deviations. 
Survey responses were examined deductively in a question-driven approach that characterized the role of generative scaffolding (\textit{e.g.,} ``...guide me through the key characters and plot points'' [P11]) and reported disagreements the participant had with the system (\textit{e.g.,} ``It was telling me to do things that weren't options'' [P09]) after each phase of the system. The interview data were transcribed automatically using Otter AI\footnote{\url{https://otter.ai}}, and were analyzed using deductive thematic analysis~\cite{braun2012thematic} to capture participants' attitudes and opinions toward \systemname. One author first independently approached the interview data, coding data that conveyed the participants' attitudes and opinions toward \systemname, and a second author revised the codes, confirming that the codes accurately characterized the participants' views.

\subsection{Findings}

\begin{figure*}[t]
    \centering
    \includegraphics[width=\linewidth]{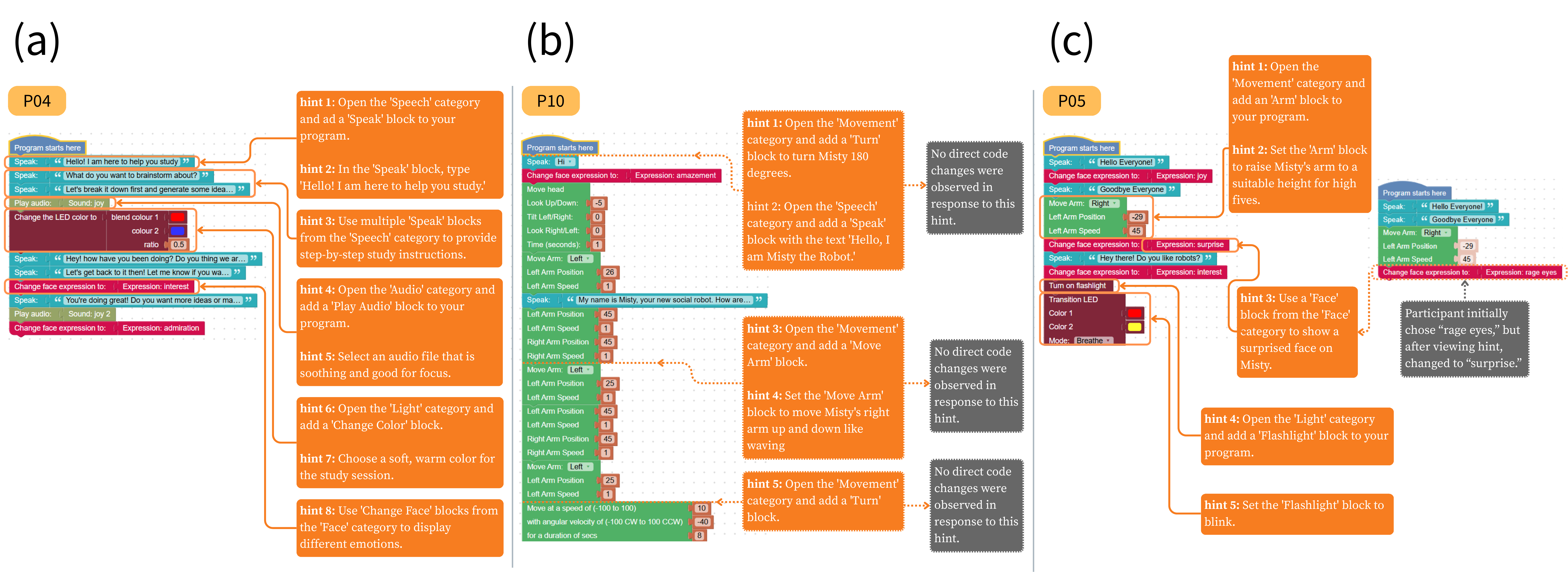}
    \caption{Three examples of use patterns observed during the programming phase. (a) Dependent: P04 actively followed and utilized all hints. (b) Independent: P10 did not follow the hints and their code remained unchanged. (c) Collaborative: P05 modified the program after receiving hint suggestions.}
    \Description{
    A figure showing three annotated examples of participant programs from the programming phase, each illustrating a different use pattern. Example (a) on the left shows participant P04, labeled as Dependent. The program is displayed as a sequence of Blockly blocks. Annotations in colored boxes show the hints P04 received and followed, including Hint 1 suggesting opening the Speech category and adding a Speak block, and Hint 2 suggesting typing specific speech text. Arrows connect each hint to the corresponding block in the program, showing that P04 directly implemented each hint in sequence throughout the entire program. Example (b) in the center shows participant P10, labeled as Independent. The program is displayed as a sequence of Blockly blocks. Annotations show hints that were provided, including hints suggesting Movement and Speech blocks. Each hint is connected to the program with an arrow and a label reading. No direct code changes were observed in response to this hint, indicating P10 reviewed the hints but did not modify their program in response to them. Example (c) on the right shows participant P05, labeled as Collaborative. The program is displayed as a sequence of Blockly blocks. Annotations show hints that were provided, including hints suggesting Movement and Light blocks. One annotation highlights a specific revision P05 made after viewing a hint, noting that the participant initially chose rage eyes as a face expression but changed it to surprise after viewing the hint suggestion, demonstrating selective adoption of scaffolding.
    }
    \label{fig:use-patterns}
\end{figure*}



Overall, our observations, interviews, and surveys revealed dichotomous views of \systemname, highlighting the strengths of \systemname and generative scaffolding, hereinafter referred to as scaffolding, to support end-user design and programming; it also brought out varying levels of frustration from participants when either \systemname or the robot's capabilities did not quite meet their expectations. 

We generally observed participants using \systemname during narrative creation to provide additional guidance for abstract details for the storyline, such as \textit{actions}, \textit{events}, \textit{endings}, and \textit{emotions}.
When programming, we observed some participants referring to scaffolding when first starting a programming task for assistance on how to start the next robot action, and when they felt their program did not execute as they envisioned.

In this section, we present our observations on how scaffolding affected design and programming, emerging use patterns of \systemname, and participants' perception of \systemname through the results of the USE and SUS questionnaire, as well as themes from our thematic analysis.

\paragraph{Interactions and Use Patterns with Generative Scaffolding}

In the narrative creation phase, five out of 14 participants requested help from \systemname at least once, with an average of 0.74 uses per participant ($SD = 1.38$; $range = 0$-$5$).
In the programming phase, 11 out of 14 participants used the hints with an average of 4.85 uses per participant ($SD = 3.57$; $range = 0$-$12$). 
A few participants noted that some tasks were missing hints when programming.

Based on our analysis of the context, character, and action changes from the storyboard to the final story summary from the narrative creation phase, we observed three use patterns emerge: independent, dependent, and collaborative. 
We categorized participants who did not request for help or requested for help but did not generally incorporate or rely on suggestions from \systemname as \textit{independent}.
We categorized participants who distinctly relied on help from \systemname to create narratives as \textit{dependent}.
In between independent and dependent, we categorized participants who referred to help, either by modifying their narrative based on help suggestions or jointly creating narratives with \systemname by responding to questions about narrative details, and demonstrated independently creating portions of the narrative as \textit{collaborative}. 

We identified eight participants (P02, P05--P07, P09--P12) as independent. 
We observed that seven out of eight independent participants did not request help from \systemname through the help button.
The remaining participant, P06, requested for help but ultimately did not accept the suggestion from \systemname and independently developed a new narrative.
We identified P14 as dependent due to his distinct reliance on \systemname to create a completely new narrative (\textit{i.e.}, context, character, actions) independent of his original storyboard and minimal original input.
We identified five participants (P01, P03, P04, P08, and P13) as collaborative.
Three participants (P01, P03, and P04) requested for help and accepted over a majority of suggestions (accepting over 77\% of suggestions) from \systemname.
Two participants (P08 and P13) did not request help via the help button; however, after further analysis of their chat interaction with \systemname, we observed the system prompting them with questions to facilitate narrative ideation, similar to the help support, and thus categorized their use pattern as collaborative.

\paragraph{Usage Patterns while Programming}

Overall, we observed participants engaging with \systemname, specifically the hints in the programming phase, in a variety of ways.
In total, 11/14 participants used at least one hint. 
Eight participants opened multiple hints in succession before programming a task.
Four participants were unable to program a task due to LLM hallucination in the provided hint (\textit{e.g.}, setting an hourly alarm---a function outside of Misty's current capabilities).
Two participants regenerated program tasks due to the inability to create a program due to a lack of robot functionality. 

During the analysis of programming, we identified three use patterns when using \systemname: dependent, independent, and collaborative.
We categorized participants who distinctly relied on hints to program as \textit{dependent}.
We categorized participants who did not use hints or referred to hints but did not make any modifications to the final program related to the hint as \textit{independent}.
In between independent and dependent, we categorized participants who referred to hints, created or modified their program based on hints, and demonstrated independently coding portions of the program as \textit{collaborative}. 
We show a selection of annotated final programs to convey and characterize observed use patterns.
The full version of all annotated programs and use pattern table with observation notes can be found in the supplementary materials. 

We identified two participants (P04 and P14) as dependent users.
We observed that both users first viewed hints, occasionally viewing more than one hint consecutively, before programming. 
For example, P04 first referred to hints, used suggested blocks by the hints, and repeated this cycle for the entirety of his program as indicated in Figure~\ref{fig:use-patterns} (a).
We identified four participants (P08, P10, P11, and P13) as independent users; a majority (3/4) did not use hints during the programming phase.
In the remaining case, we observed distinct usage of hints from P10, where he reviewed hints; however, he ultimately did not make changes to the program in response to the hint, as seen in Figure~\ref{fig:use-patterns}(b).
We identified eight participants (P01--P03, P05--P07, P09, and P12) as collaborative, demonstrating a range of collaboration with \systemname. 
They alternated between independently programming and using hints.
We note that P09 regenerated hints due to a hallucination of robot capabilities. 
P05 referred to hints to identify what category of block to choose from (\textit{e.g.}, selecting a block from the `Movement' category) and made revisions to portions of her own code after viewing a hint (\textit{e.g.}, changing a face expression from `rage eyes' to `surprise') as seen in Figure~\ref{fig:use-patterns}(c).

\paragraph{Comparison of Storyboard to Robot Program}

Before using the system \systemname, we asked participants to create a storyboard depicting the human-robot interaction they would later program. 
Then, during the interview, we asked participants to review the storyboard they initially created and compare it to the final program they deployed on the Misty robot to reflect on their experience working with and without scaffolding.
Participants shared diverse thoughts, some crediting the creative input from the LLM support provided to them to transform their stories, while others felt their story remained relatively consistent.
We identified nine out of 14 participants whose story summary included embellishment in narrative details from \systemname (\textit{e.g.}, adding details to the background environment) without modifying context, character, or actions.
Participants who felt the scaffolding and LLM positively affected their narrative outcomes shared reasons such as adding descriptions and details to actions (P01, P03, P12), filling in blanks (P02), providing fun suggestions (P03, P06), and visualizing the story (P04).
Others noted minor changes in alignment with the robot capabilities suggested by the scaffolding (P05).
P07 explained that scaffolding did not significantly change their story, while others shared they felt their story was pushed by scaffolding in a direction that did not align with the experiences of the Misty robot's capabilities (P8, P10, P13).

\paragraph{Participant Perception of Generative Scaffolding}

\begin{figure}
    \centering
    \includegraphics[width=3in]{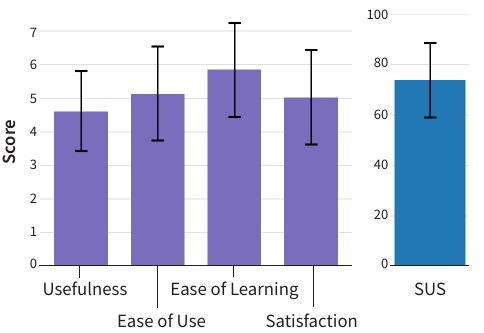}
    \caption{Mean and standard deviations of user perceptions of \systemname across the USE subscales (usefulness, ease of use, ease of learning, satisfaction) and the SUS.}
    \Description{
    A bar chart showing mean scores and standard deviation error bars for user perceptions of the system across five measures. The horizontal axis lists five measures: Usefulness, Ease of Use, Ease of Learning, and Satisfaction from the USE questionnaire, and SUS from the System Usability Scale. The chart uses two vertical axes: the left axis shows USE scores ranging from 0 to 7, and the right axis shows SUS scores ranging from 0 to 100. The four USE subscale bars are shown in purple on the left scale. Usefulness shows a mean of 4.62, Ease of Use shows a mean of 5.14, Ease of Learning shows a mean of 5.85, and Satisfaction shows a mean of 5.03. The SUS bar is shown in purple on the right scale with a mean of 73.85. All bars include error bars indicating standard deviations. Ease of Learning shows the highest USE score among the four subscales, while SUS falls in the acceptable range on the 100 point scale.
    }
    \label{fig:use-sus-results}
\end{figure}

As one of the participant's responses (P05) was not properly recorded, we report the SUS score of 13 participants ($M = 73.85$, $SD = 14.78$) and USE subscales of perceived usefulness ($M = 4.62$, $SD = 1.19$), ease of use ($M = 5.14$, $SD = 1.40$), ease of learning ($M = 5.85$, $SD = 1.41$), and satisfaction ($M =5.03$, $SD = 1.46$), as shown in Figure~\ref{fig:use-sus-results}.
We found the SUS score and USE subscales generally aligned with the participants' remarks shared via survey responses and interviews. 

We asked each participant to describe the role of scaffolding at each phase of \systemname: story creation, planning, and programming. 
Below, we summarize their thoughts and level of agreement with the responses generated by \systemname. 
When reflecting on their experience with \systemname during the story creation efforts, participants praised that it was overall ``helpful'' (P01, P02, P04, P07, P08, P12--14) and noted its ability to clarify ideas (P03, P05), and keep track of thoughts (P03, P08, P12).
P10 characterized the scaffolding as ``inquisitive'' and perceived its behavior as ``want(ing) data.''
During this phase, no participants reported any disagreement with the scaffolding. 

Following the story creation, the participants' response to \systemname during the planning phase was similar, echoing praise of being ``helpful'' (P01, P02, P08, P12, P14). 
Participants elaborated that  \systemname helped instruct participants on what to do next (P01, P12), simplify concepts (P02, P12), and summarize the key plots (P08). 
Other participants noted the scaffolding from \systemname connected to the next phase of programming (P03, P05, P06, P09, P11, P12), sharing that they felt it provided ``guidelines'' (P06) and ``made it easier to visualize the timing and sequence'' (P03).
During this phase, P09 reported a disagreement with \systemname, stating: ``[\systemname] was telling me to do things that weren't options. I wanted Misty to go off every hour, but there was not a timer block like the [scaffolding] said there would be.'' 
This resulted in one out of 14 participants reporting disagreement between the participants and the scaffolding suggestions.

Finally, during the programming phase, we observed more divergent opinions toward the \systemname from the participants. 
While some participants noted the usefulness of the hints (P01, P03, P05, P06, P09, P12, P14), others shared they had limited interaction with the \systemname (P02) or felt scaffolding in general was not required at this phase (P07, P10).
Participants expressed having conflict as some of the suggestions did not properly reflect the robot's capabilities (P05, P09) or were missing hints (P06).
P13 mentioned she ``wanted more than what was on the checklist [of programming tasks].''
This resulted in four out of 14 participants reporting disagreement between themselves and the scaffolding support. 

\paragraph{Views on Narrative Creation and Robot Programming}

Participants had various perspectives on narrative creation and robot programming. 
P01 shared that it was her favorite part of the process: ``I like more the story part... because you really can't get anywhere without a story.''
P04 expressed how they felt the storyboarding and later narrative creation added to his experience: ``I feel like the storyboarding part really added to that process, where I could actually visualize it first on paper and have some rough sketches around it and create a story, and then have that story made into, like a summary.''

However, not all participants felt the narrative creation and related scaffolding were necessary, and shared their reflections.
P06 articulated her positive thoughts and contemplated how others would view integrating narratives: ``I think it definitely made me more excited to program it... I thought it was cool to come up with a fun scenario, which I think, for new programmers, would be cool to get people excited to want to work with it, and get to be creative. But yeah, I think not everyone would want to do that [narrative creation] I guess... I know a lot of people who would not want the story.''
P02 noted his complex stance on integrating narratives and scaffolding as it provided support and also led him astray: ``[Narrative creation] really did help fill in the blanks, because I didn't know what I was doing, I didn't know what you wanted me to do, and I didn't know what kind of story I wanted to tell, and the blocking out did help with that. [Narrative creation] kind of called for more details than were actually going to be used, I guess, for better [or] for worse... But I guess it also led me astray, not understanding what the limits of the programming with the robot could actually do.''

\paragraph{Juxtaposition of Scaffolding and Robot Capabilities}
During the narrative creation phase, the first phase for end-user design and programming, participants praised \systemname's ability to intuitively support planning and later translate the narrative into actionable programming tasks. 
However, during the programming phase, participants expressed frustration while trying to browse through the different available programming blocks.
At times, the participants felt they were presented with conflicting information, as the scaffolding gave a suggestion incompatible with the available robot capabilities in programming blocks.

P02 voiced his thoughts on the limited capacity of the robot during the programming exercise: ``I wasn't interested in that level of detail [or programming], and because it had so little capability, had no capability to understand its environment... I was expecting more from a modern [robot], and it's not a criticism of the programming interface, it's just a criticism of what [the robot] can do. It'll evolve to control some of those other things in the future... Knowing what it can do, it's a robot, and uses much more AI and sensors in understanding its environment...''
P02 further suggested a solution emphasizing transparency: ``The most important thing would be to say it doesn't know its environment. And from that, a lot more understanding would take place. You're really programming it to do things. It doesn't interact with its environment...''
P09 debated about the tradeoffs of knowing more about the robot's capabilities: ``The story changed because realizing... there [were] going to be some limitations of things that I thought that I didn't know [the robot] couldn't do while I was creating the storyline... sometimes it's good to not have those stipulations [about capabilities] and just be creative. And when you're brainstorming too, what is your ideal goal... this is kind of like a double-edged sword.''
To better convey the robot's abilities, a few participants suggested presenting a list of capabilities or preset motions (P02, P05) or having visual previews (P12) on the interface to increase familiarity with available robot actions. 

\section{Discussion}

Through our empirical exploration, we sought to understand how participants use generative scaffolding for end-user robot programming within our designed \systemname interface. 
We thus review our findings and discuss the research questions in this section. 

\paragraph{(RQ1) How can generative scaffolding support end-user design and programming of social robot behaviors?}


Generative scaffolding was a double-edged sword, providing support for some participants, while others felt the scaffolding introduced distraction. 
We identified several instances where \systemname contributed suggestions that were reflected in the final narrative.
Our findings confirm that generative scaffolding contributed to creativity in planning and narrative development, as demonstrated in applications of generative AI to support creativity~\cite{Hwang2025-humanaidigitalcontentmaking}. 
We observed instances where scaffolding proactively generated ideas by proposing new context or characters for the participant to build upon.
In addition to creative input, we also observed scaffolding to provide embellishing details during planning. 
For example, participants shared that scaffolding helped flesh out character details, deepened descriptions of background environments, and added details for programmable actions. 

However, not all scaffolding contributed positively to the design and programming process.
Participants reported moments when the generative scaffolding hallucinated and introduced confusion.
While unintentional, participants reported exhaustively going through folders of features to find a programming block that matched the suggestion.
These instances distracted participants from properly executing the planning and programming tasks, as the participant was fixated on the hint browsing Misty's functions for something that does not exist.
To address this issue, scaffolding should dynamically adjust to users' needs, supporting areas where they perceive gaps. 

\paragraph{(RQ2.) What interactions, use patterns, and user perceptions emerge when using generative scaffolding?}

The study revealed distinct interactions and use patterns that illustrate how novices engaged with generative scaffolding.
While our analysis yielded three use pattern categories, triangulation across interview and observational data revealed additional nuance not fully captured by reliance patterns alone. Within the collaborative category, we observed distinct engagement styles, and scaffolding-capability mismatches produced a qualitatively different experience that we address as a distinct persona. We therefore present four personas: three capturing variation along the reliance spectrum from independent to dependent use, and one capturing the distinct experience of frustration arising from mismatches between scaffolding suggestions and available robot capabilities.
We characterize these use patterns with user personas to highlight a range of reliance on scaffolding, variation in strategy and action, and points of friction: 

\textbf{Iyori, the Independent Builder} -- Iyori exemplifies independence. 
She does not request help during narrative creation and uses no hints or scaffolding while programming. 
Her narrative shows only minor embellishments, and her story elements of characters, context, and robot actions remain consistent in both her storyboard and program. 
This pattern underscores that some novices prefer to maintain control, agency, and continuity, relying minimally on system scaffolds.

\textbf{Daku, the Dependent} -- Daku reflects a heavily dependent user, reliant on scaffolding. 
He begins creating a narrative by repeatedly selecting the ``Help Me'' button before writing, and continues this strategy in the programming phase, first using hints step by step to assemble his program. 
This pattern highlights how scaffolding can become the \textit{primary driver} of user progress, raising questions about exercising user agency and following guidance provided by scaffolding.

\textbf{Franka, the Frustrated} -- Franka exhibits collaborative tendencies, co-creating a compelling narrative with \systemname about Misty playing music for a party. 
The scaffolding suggests she selects a song for the party scenario.
Yet, when her programming library and environment does not have the desired resources (\textit{e.g.}, songs in the `Audio' category), she becomes stuck and frustrated, spending considerable time searching each programming category before abandoning the task. 
This case illustrates how mismatches between user intent and robot capability can undermine engagement and trust in scaffolding.

\textbf{Harry, the Hint Collector} -- Harry demonstrates a hybrid reliance between dependent and collaborative users.
While creating the narrative, he mostly accepts suggestions on the direction of the story.
Then, before programming, he exhaustively reviews all available hints, using them as a source of inspiration and guidance for the next programming steps. 
He creates a program with few revisions to the generated narrative, and that minimally strays from the scaffolding.  
This pattern suggests that some novices are strongly influenced by the scaffolding and treat system-generated hints as a form of \textit{scaffolding library}, a way to gain knowledge and orient themselves before committing to programming.

Together, these personas highlight a spectrum of novice engagement with generative scaffolding, ranging from independent to dependent use, and from productive use to frustration. 
This suggests that while scaffolding should not only lower barriers to entry and support creativity, but also align with user expectations, clearly communicate robot capabilities, and actively help bridge gaps between narrative goals and executable robot behaviors.

\subsection{Design Insights}
Based on our analysis of how novice users engaged with generative scaffolding across four phases, we distill three design insights. These are for end-user design and programming environments for social robots that seek to integrate generative AI as a scaffold. 
While our findings are grounded in a linear, narrative-based robot programming workflow, the tensions we observed between guidance and agency, suggestion and feasibility, and progression and continuity point to broader considerations for generative scaffolding in more complex interactive systems. Social robots capable of iterative, reactive, or state-dependent behaviors would introduce additional complexity where these tensions would be amplified. Our insights therefore extend beyond our current linear pipeline to inform the design of generative scaffolding in such settings.

\paragraph{Design Insight 1: Generative scaffolding should proactively support users in calibrating their understanding of the robot's capabilities throughout the narrative creation and planning process.} 
In \systemname, both the interface and LLM prompts explicitly described the range of robot functionality and were always visible in the narrative creation and planning phases. 
Despite this, novices created narratives that did not align with the range of robot capabilities, and in some cases the scaffolding itself hallucinated and provided suggestions that conflicted with the robot's available capabilities. 
This highlights a tension in generative scaffolding, while it supports ideation and expression, it can also amplify mismatches between user intent and system constraints.

These observations suggest that scaffolding should not only support ideation but also actively engage users to examine whether their narrative aligns with the robot's capabilities. 
Just as it is essential to maintain users' understanding of the programming loop to safeguard against forgetfulness~\cite{chen2023forgetful}, it is essential to uphold transparency and continuously confirm and update expectations throughout the process to ensure the outcome aligns with expectations \cite{ajaykumar2021survey, weintrop2018evaluating, 10.1145/1640233.1640240, 10.1145/3267782.3267921}. 
Such support can be achieved by reflective nudges, prompting users to verify whether elements of their narrative align with known capabilities or warnings that surface potential mismatches while preserving user agency to proceed or revise.
In our study, this calibration challenge persisted even with explicit capability descriptions visible on the interface. In more complex interactive systems, a user might design a scenario where the robot responds differently depending on who it detects or what it hears, only to discover during programming that the robot lacks sensing capabilities to support those interactions. Generative scaffolding should therefore surface capability constraints not just at the start of narrative creation, but each time the user introduces a new robot behavior or response condition.

Future work could examine alternative representations of robot capabilities to ensure better alignment between robot capabilities and user expectations. 
This includes the exploration of different visualization techniques, such as animated story previews, contextual capability indicators, or interaction histories that expose robot limitations at appropriate stages in the pipeline and storage of useful hints from past tasks.

\paragraph{Design Insight 2: Generative scaffolding should dynamically adjust to end users' needs, supporting areas where they perceive gaps while enabling them to update and redefine the goals of scaffolding.}
Design and programming are inherently iterative activities, involving ongoing refinement during ideation, reflection, and refinement.
In our study, generative scaffolding in \tool was grounded in a prompt globally used across participants.
For some participants, the scaffolding provided suitable and creative feedback; for others, the scaffolding seemed irrelevant or distracting as they wanted more support to confirm their narratives aligned with the robot's capabilities.
These observations suggest that fixed scaffolding strategies may be insufficient for supporting diverse expectations, expertise levels, and intentions among novice users.
Instead, effective generative scaffolding should be adaptive or reflexive, allowing its form and focus to shift overtime. Such adaptability may involve enabling users to redefine what kind of support they want, or allowing the system to adjust its scaffolding in response to interaction patterns, task context, or mismatches that emerged between narrative intent and executable robot behavior. 
In non-linear applications, the user may need to navigate branching task paths, revisit earlier programming decisions, and reconcile multiple intermediate states. To support this complexity, generative scaffolding should enable modular interaction structures, where individual tasks or suggestions are organized into separable units. This would allow users to selectively revise or regenerate misleading suggestions without disrupting other parts of their workflow, supporting more flexible navigation across branching or iterative task paths.

Future work could explore flexible scaffolding designs that begin with a shared baseline but allow users to revise prompts, update the type of feedback they receive, or change the balance between ideation and verification as their understanding evolves.

\paragraph{Design Insight 3: Programming environments that adopt a generative scaffolding approach should clearly demonstrate connections between narrative creation and programming outcomes.} 
While many participants expressed their appreciation for the full-pipeline process from narrative creation to deployment, some participants felt uncertain during the process about how their current efforts would lead to the final program. This uncertainty highlights the importance of reinforcing continuity across phases, particularly when moving from expressive design activities to logical programming tasks.
To support the transition from intent to robot programming, environments that adopt generative scaffolding should clearly demonstrate how narrative elements are transformed into goals and how those goals shape executable robot behavior. Making these connections visible helps users understand that each phase is not an isolated exercise, but a step toward realizing their intent.
This gap would likely widen in a non-linear application where breakdowns in earlier stages would propagate across branching interaction paths, making it difficult to trace how specific narrative elements map to programmable actions. To support users in understanding the progression and dependencies, generative scaffolding systems should make relationships between steps explicit. 
In linear workflows, this may take the form of sequential representations (\textit{e.g.,} timelines), while in non-linear workflows, more expressive structures, such as tree diagrams or network graphs, can help users trace progress, explore alternatives, and diagnose breakdowns across interconnected tasks.

Future work could explore designing visual connections within the interface or incorporating lesson-based plans to support continuity across phases. For example, \citet{10.1145/3654777.3676357} presents methods for facilitating hierarchical task decomposition and enabling direct manipulation of code segments.

\subsection{Limitations and Future Work}
Our work has a number of limitations. 
First, during the study, the interactive capabilities of the robot were constrained. 
Although, the Misty robot supports autonomous interaction through its numerous sensors and buttons, these functions were not integrated into \systemname. This design choice was intentional, allowing us to focus on how generative scaffolding supports narrative-based planning and goal-driven programming without introducing additional sources of complexity. As a result, our findings primarily reflect interaction design scenarios where robot behavior is scripted rather than dynamically reactive. More broadly, our findings reflect scripted social robot behaviors that does not account for the reactive and adaptive behaviors central to social robotics~\cite{BREAZEAL2003167}. However, the deployment to a physical robot was not incidental. Participants programmed expressive social behaviors including gestures, facial expressions, and speech. Also, their expectations of robot capabilities and their reactions to seeing programs execute on a physical platform reflect robotics aspect of the experience. Programming a truly social robot requires reasoning not only about what the robot should do, but how it should adapt in response to a person's actions and social cues in the moment. This is a dimension our current approach does not yet address.
Expanding \systemname to support richer interaction paradigms, such as trigger-action programming or sensor-driven behaviors, represents an important direction for future work.

Second, our empirical exploration approach also presents limitations. 
We acknowledge that our interviews were relatively brief (5--16 minutes), which may have limited the depth of qualitative insight into participants' attitudes, agency, and perceptions of generative scaffolding. The interviews were designed to capture immediate reflections rather than extended retrospective narratives. To mitigate this limitation, we complemented interview data with qualitative survey responses, usage logs, and observation notes, triangulating across multiple data sources to capture both in-the-moment behaviors and participant interpretations. We recognize that future work employing longer interviews or diary studies could provide richer insight into how novices develop attitudes toward generative scaffolding over time. Beyond interview depth, our research scope also limited our ability to directly compare narrative-based programming with alternative approaches such as traditional block-based programming or LLM-based programming.
Future work could build on our findings through comparative or ablation studies that systematically vary scaffolding components or contrast narrative-based workflows with other programming approaches to further understand the trade-offs involved. 

\section{Conclusion}
In this paper, we proposed a novel four-phase narrative programming approach for using narratives to facilitate novice end-user design and robot programming. 
Through a Research through Design process, we designed and developed the \systemname interface to facilitate this approach, and understand how generative scaffolding can support novice social-robot programmers. 
Our findings demonstrated the strengths of generative scaffolding, and highlighted distinct user personas and usage patterns that illustrate how scaffolding shapes end-user design and programming strategies.
Based on our findings, we presented design insights that emphasize alignment between end-user intent and executable behaviors and suggested future directions for generative scaffolding in end-user design and end-user robot programming for social robots.

\begin{acks}
We thank all our participants for sharing their time and insights.
This work was supported by the Sheldon B. and Marianne S. Lubar Professorship and the National Science Foundation award (\#1925043).
\end{acks}

\bibliographystyle{ACM-Reference-Format}
\bibliography{references}

\end{document}